# *The Solar Spectroscopy Explorer Mission*

Jay Bookbinder[1]§, J. Adams[3], D. Alexander[8], M. Aschwanden[12], C. Bailey[3], S. Bandler[3], S. Bradshaw[8], N. Brickhouse[1], J. Chervenak[3], S. Christe[3], J. Cirtain[3], S. Cranmer[1], S. Deiker[12], E. DeLuca[1], G. Del Zanna[13], B. Dennis[3], G. Doschek[2], M. Eckart[3], F. Finkbeiner[3], A. Fludra[15], P. Grigis[1], R. Harrison[15], L. Ji[1], C. Kankelborg[14], V. Kashyap[1], D. Kelly[3], R. Kelley[3], C. Kilbourne[3], J. Klimchuk[3], Y.-K. Ko[2], J.M. Laming[2], E..Landi[6], M. Linton[2], D. Longcope[14], J. Mariska[2], D. Martinez-Galarce[12], H. Mason[13], D. McKenzie[14], R. Osten[17], G. Peres[10], A. Pevtsov[16], K. Phillips[4], F. S.Porter[3], D. Rabin[3], C. Rakowski[2], J. Raymond[1], F. Reale[10], K. Reeves[1], J. Sadleir[3], D. Savin[11], J. Schmelz[9], R.K. Smith[1], S. Smith[3], R. Stern[12], J. Sylwester[5], D. Tripathi[13], I. Ugarte-Urra[7], P. Young[7], H. Warren[2], B. Wood[2]

[1]Smithsonian Astrophysical Observatory, [2]Naval Research Laboratory, [3]NASA Goddard Space Flight Center, [4]Mullard Space Science Laboratory, UCL, [5]Space Research Centre, Polish Academy of Sciences, [6]University of Michigan, [7]George Mason University, [8]Rice University, [9]University of Memphis, [10]University of Palermo, [11]Columbia University, [12]Lockheed Martin, [13]DAMTP, University of Cambridge, [14]Montana State University, [15]STFC Rutherford Appleton Laboratory, [16]National Solar Observatory, [17]Space Telescope Science Institute

## EXECUTIVE SUMMARY

The Solar Spectroscopy Explorer (SSE) concept is conceived as a scalable mission, with two to four instruments and a strong focus on coronal spectroscopy. In its core configuration it is a small strategic mission ($250-500M) built around a microcalorimeter (an imaging X-ray spectrometer) and a high spatial resolution (0.2 arcsec) EUV imager. SSE puts a strong focus on the plasma spectroscopy, balanced with high resolution imaging – providing for break-through imaging science as well as providing the necessary context for the spectroscopy suite. Even in its smallest configuration SSE provides observatory class science, with significant science contributions ranging from basic plasma and radiative processes to the onset of space weather events. The basic configuration can carry an expanded instrument suite with the addition of a hard X-ray imaging spectrometer and/or a high spectral resolution EUV instrument – significantly expanding the science capabilities. In this configuration, it will fall at the small end of the medium class missions, and is described below as SSE+. This scalable mission in its largest configuration would have the full complement of these instruments and becomes the RAM (Reconnection And Microscale) mission. This mission has been designed to address key outstanding issues in coronal physics, and to be highly complementary to missions such as Solar Probe Plus, Solar Orbiter, and Solar-C as well as ground-based observatories.

The rationale for a scalable concept was to allow the Decadal committee to select the most appropriate size mission for additional study, once budget constraints are understood. The nominal mission design life is three years, with consumables sized for 10 years. The observatory design is built on studies performed over the last decade by SAO, GSFC, and several commercial spacecraft providers, and has strong heritage in all components from previous space flight missions. This submission is one of a trio of white papers discussing how new instrumentation (Bandler et al. 2010) can provide breakthroughs in heliophysics (Laming et al. 2010).

## 1. SUMMARY OF SCIENCE CONCEPT

The Solar Spectroscopy Explorer (SSE) addresses fundamental and timely questions in solar physics. Solar physics, like much of heliophysics, is a mature science. No single instrument can comprehensively address the key science questions outlined in the 2009 Roadmap or the 2003 Decadal Survey. SSE's scalable approach identifies the smallest instrument complement that can address a set of strategic science questions and outlines how additional instrument(s) expand the capabilities of the mission. The reader is referred to the numerous white papers in particular that discuss the breadth of science opportunities achievable via high-resolution X-ray spectra and imaging (e.g. Laming et al. 2010, Golub et al. 2010, Lin et al 2010, Smith et al. 2010, Osten et al. 2010, Kashyap et al. 2010) as well as instrument white papers on the XMS (Bandler et al., 2010b), FOXSI (Christe et al. 2010), and FACTS (Korendyke et al. 2010).

*§Contact information: jbookbinder@cfa.harvard.edu; 617-495-7058*

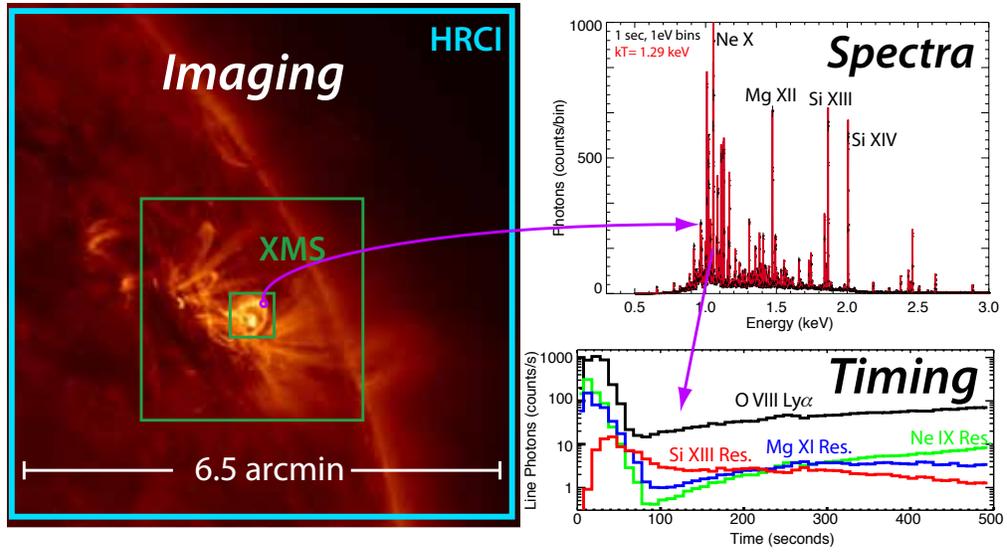

*Figure 1 –SSE combines imaging, spectra, and timing, as it can image regions at arcsecond resolution over a wide field of view, creating high-resolution spectra at each point and tracking the evolution of individual emission lines as a function of time. Shown here is a post-CME active region observed at 171Å with SDO/AIA, with the FOV of the core & extended XMS array (in green) as well as the HRCI (in blue) overlaid. A 1s spectrum from a 100 sq. arcsec extraction region is shown, along with one possible evolution of each line's flux over a 100 minute observation. The FOXSI FOV (10×10 sq. arcmin) extends beyond the image.*

    The corona radiates most of its energy in the soft X-rays, emitting resolvable lines that provide a wealth of spectral diagnostics. These soft X-ray emission lines tell the story of how the plasma is heated, where the plasma is heated, and for how long the plasma has been heated via measurements of the temperature, density, and time-dependent ion populations as well as departures from Maxwellian velocity distributions. These will be combined with measurements of flow and turbulent velocities from line profiles and line shifts. These are the measurements that are necessary to understand the physics of magnetic reconnection, coronal heating, flare and CME energetics, and solar wind acceleration (Table 1).

## 2. ADVANCING SOLAR PHYSICS IN THE NEXT DECADE

    Unlike previous missions, SSE will provide excellent imaging, spectral, and timing information simultaneously. One of the two core instruments, the X-ray Microcalorimeter Spectrometer (XMS), will enable measurements with unprecedented spectral (<2 eV) and temporal (milliseconds) resolution over a broad spectral range (0.2-10 keV). Unlike narrow-band or single-line imagers, all of the coronal plasma can be seen, enabling comprehensive studies of its evolution – critical information as the corona is dynamic, necessitating that all of the diagnostic lines be observed simultaneously for a complete picture. The high temporal resolution of the XMS will characterize flare energy distributions with unprecedented accuracy, in a manner that has been applied successfully for other stars (Kashyap et al. 2002). These are known to be distributed as a power-law over many orders of magnitude (Hudson 1991); photon counting enables sophisticated analyses (developed for low signal stellar data) that explore a variety of physics at different timescales and energy ranges.

    This X-ray spectroscopy capability is complemented by the second core instrument: the high resolution coronal imager (HRCI), a normal incidence system at the 193Å (Fe XII, XXIV; log T=6.1, 7.3) and 131Å (Fe XX; Log T=7) passbands that provides high cadence and high spatial resolution (0.1" pixels) imaging over a ~7×7 sq. arcmin FOV (~14×14 sq. arcmin goal). This is a factor of five better imaging resolution than currently available on AIA. The HRCI allows us to follow the evolution of the fine-scale coronal sub-structure, whose scale is based on spectroscopic observations that indicate that ~10% of the observed area is filled with emitting plasma (see Golub et al. 2010 and Doschek 2010). The associated topological and dynamical contributions from the HRCI are shown in Table 1. The HRCI is particularly sensitive to low amplitude transverse oscillations that may result from small impulsive events or waves that originate in the lower atmosphere.

**Table 1 – Science Flowdown for SSE & SSE+ Mission Options**

| Science | | | Observatory Requirements | | | | |
|---|---|---|---|---|---|---|---|
| Topic | Target | Physical Process | Spatial Resol. | Spectral Resolution | Timing | Measurement | Instrument |
| | | | arcsec | $\Delta E@E$ (keV) | seconds | | |
| **Magnetic Reconnection** | Active Regions | Heating | 1 | 0.002@1 | 10 | $T_e(s), n_e(s)$ | XMS |
| | | Flows | 1 | 0.002@1 | 10 | $v(T_e,s,t)$ | XMS, HRCI |
| | | Particle Acceleration | 10 | 1@15-30 | ~0.001 | $dn_e(s)/dE$ | FOXSI (Pri) |
| | | | 1 | 0.002@1 | 1 | $dn_e(s)/dE$ | XMS (Sec) |
| | | Dynamics/Topology | 0.2 | – | 1 | $v(s,t), I(s,t)$, connectivity | HRCI |
| **FIP Fractionation/ Coronal Heating** | Active Regions & Coronal Loops | Heating | 1 | 0.002@1 | 10 | $T_e(s), n_e(s)$ | XMS |
| | | Flows | 1 | 0.002@1 | 10 | $v(T_e,s,t)$ | XMS, HRCI |
| | | Dynamics/Topology | 0.2 | – | 2 | $v(s,t), I(s,t)$, connectivity | HRCI |
| | | Abundances | 1 | 0.002@1 | 10 | Ion fractions | XMS |
| **CME Energetics** | Behind-/Off- Limb Flares | Heating | 5 | 0.01@6 | 5 | $T_e(s), n_e(s)$ | XMS |
| | | Dynamics/Topology | 0.2 | – | 1 | $v(s,t), I(s,t)$ | HRCI |
| **Solar Flares** | Flaring Active Regions | Heating | 1 | 0.002@6 | 0.1 | $T_e(s), n_e(s)$ | XMS |
| | | Flows | 1 | 0.002@6 | 10 | $v(T_e,s,t)$ | XMS, HRCI |
| | | Particle Acceleration | 10 | 1@15-30 | ~0.001 | $dn_e(s)/dE$ | FOXSI (Pri) |
| | | | 1 | 0.002@6 | 1 | $dn_e(s)/dE$ | XMS (Sec) |
| | | Dynamics/Topology | 0.2 | – | 1 | $v(s,t), I(s,t)$, connectivity | HRCI |
| **Solar Wind Acceleration** | Off-Limb Coronal Holes | Heating | 1 | 0.002@1 | 30 | $T_e(s), n_e(s)$ | XMS |
| | | Flows | 1 | 0.002@1 | 30 | $v(T_e,s,t)$ | XMS, HRCI |
| | | Dynamics | 0.2 | – | 1 | $v(s,t), I(s,t)$ | HRCI |
| | | Abundances | 1 | 0.002@1 | 10 | Ion fractions | XMS |

*Notes: $T_e(s)$ & $n_e(s)$ are electron temperature & density as a function of position, while $v(T_e,s,t)$ is the velocity as a function of electron temperature and/or position & time. $dn_e(s)/dE$ is the electron energy distribution; this can be measured with FOXSI directly, or indirectly with spectra from XMS. $I(s,t)$ is the EUV intensity.*

This core mission can be extended in a number of ways. The most natural addition is a hard X-ray imaging spectrometer (FOXSI; see Christe et al. 2009), whose focusing optic allows for more detailed studies of the non-thermal electron population. The energetic contribution of non-thermal particles in reconnection events tightly constrains models of the reconnection process. The combination of the XMS (which follows the influence of the non-thermal electrons on the ion populations and indirectly derives their distribution) with FOXSI (which directly images the non-thermal electrons and thus their distribution functions) provides a unique capability for understanding coronal heating, electron acceleration, and CME acceleration. This three instrument configuration (XMS, HRCI, FOXSI) is termed SSE+. An alternative SSE+ implementation would replace FOXSI with an EUV spectrometer (such as FACTS; Korendyke et al. 2010) which provides several exciting capabilities: spectroscopic analysis of transition regions and low coronal lines, better velocity resolution (~5 km/s). These capabilities would require rastering a slit across the FOV and so sacrifices the time resolution needed to follow rapid events.

Significant science contributions from SSE range from understanding basic plasma and radiative processes in the corona to providing fundamental measurements of the onset of space weather events. Working together, the instruments comprising SSE/SSE+ will allow us to investigate the propagation of flare and CME disturbances throughout the solar atmosphere, and pinpoint the location and height of magnetic reconnection, energy deposition, and solar wind acceleration.

## 3. IMPLEMENTATION

The core SSE mission is conceived as a small strategic mission (i.e., with a cost in the range of FY10 $250M - $500M) that will be placed into a geosynchronous orbit using either an Atlas-V or a Falcon 9. The LV selection is driven by the weight of the calorimeter and requirement for a geosynchronous orbit (set after trade studies that were driven by data rates). As a result, there is significant excess launch capacity that can be exploited by augmenting the instrument suite and moving the mission to the low-cost end of the medium-sized category (SSE+). We describe both implementations below.

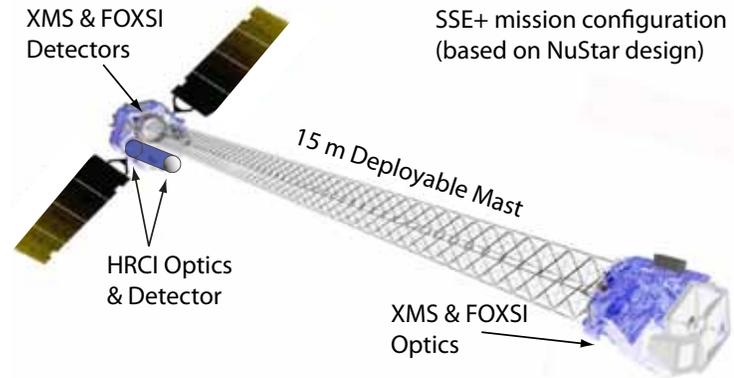

*Figure 2 – Schematic of the SSE+ mission; core SSE omits FOXSI.*

### 3.1 SSE Instrument Module

The SSE payload (see cartoon in Fig. 2) consists of (1) a grazing incidence (GI) Flight Mirror Assembly (FMA) mounted on the end of the extendable optical bench (EOB) module; (2) the XMS detector mounted at the focal plane; and (3) the HRCI that consists of a co-aligned self-contained AIA-like telescope.

The FMA design provides effective area of ~2 cm² at 1.25 keV, and ~1.5 cm² at 6 keV. The XMS (Bandler et al. 2010a) provides high spectral resolution (~2 eV over the 0.2-10 keV bandpass), non-dispersive imaging spectroscopy over the central array with 1 arcsec resolution and a field of view of 36×36 sq. arcsec. An extended array expands the FOV to 3×3 sq. arcmin over the same bandpass with slightly lower energy resolution (≤5 eV). Since the devices are photon counting, each X-ray photon arrival is time-tagged with microsecond precision. Count rates up to 1000 cps can be accommodated with a slight degradation in spectral resolution. The dynamic range of the XMS is increased to over 100,000 with a filter wheel such as used on the Hinode XRT. A adiabatic demagnetization refrigerator (ADR) and a mechanical cryocooler provide cooling to 50 mK without expendable cryogen – so there is no intrinsic limitation to the mission lifetime.

The HRCI is a higher-resolution version of an AIA telescope consisting of a 0.4m diameter Ritchey-Crétien optic that – similar to AIA – is coated in segments to allow for multiple EUV channels (nominal wavelengths are: 194Å and 131Å). As with TRACE and AIA, a guide telescope provides the image stabilization signal to a steering secondary optic. The detector for HRCI is a 4k×4k (8k×8k goal to increase the FOV) CCD. The most significant change from AIA is the plate scale (0.1 arcsec instead of 0.6 arcsec pixels). The expected cadence of the HRCI is ~1s.

### Table 2 – Resource Summary for SSE & SSE+

| Payload | Mass (kg, CBE) | Power (W, CBE) | Data Rate (mbps) |
|---|---|---|---|
| XMS | 300 | 650 | 22.8 |
| HRCI | 50 | 50 | 22.4 |
| S/C (incl. EOB) | 350 | 150 | 1 |
| Total CBE | 700 | 850 | 46.2 |
| Reserves (30%) | 210 | 255 | 13.9 |
| SSE Mission Totals | 910 | 1105 | 60 |
| S/C capability | | 1500 | 130 |
| LV throw to GEO | 2000 | | |
| **Margins** | **55%** | **26%** | **54%** |
| FOXSI | 100 | 60 | 1 |
| SSE+ CBE | 800 | 910 | 47.2 |
| SSE+ Reserves | 240 | 273 | 14.2 |
| SSE+ Mission Totals | 1040 | 1183 | 61.4 |
| **SSE+ Margins** | **48%** | **21%** | **53%** |

### 3.2 Extendable Optical Bench (EOB) Module

The EOB is the portion of the metering structure which is extended on orbit. It consists of an ADAM mast, almost identical to that on the NuSTAR SMEX mission. High precision deployment accuracy/repeatability was proven with the 60 m ADAM used in space on the NASA's Shuttle Radar Topography Mission, allowing the XMS and FOXSI to have very modest focus mechanisms. A simple metrology system (RHESSI or NuSTAR heritage) will track and correct motions for the two instruments that use the EOB – FOXSI and XMS – to enable accurate attitude reconstruction.

### 3.3 SSE+ Instrument Module

The addition of a hard X-ray imager is included in the SSE+ payload; the FOXSI instrument is used as a specific example. FOXSI extends the SSE+ mission response to ≥60 keV with 8 arcsec FWHM using nested grazing incidence optics and depth-graded multilayers to provide ~180 cm$^2$ of effective area at 40 keV; the FOXSI optic has the same focal length as the XMS optic, and is mounted to the same EOB. The FOXSI instrument uses 256×256 CdTe or CdZnTe detectors that provide 500 micron (6 arcsec) pixels, an energy resolution of 1 keV from 5-60 keV, and a FOV of 10×10 sq. arcmin (limited by optics vignetting). Background suppression is achieved with an active anticoincidence shield surrounding five sides of the imager, with heritage from NuSTAR and Astro-H.

### 3.4 Spacecraft Module

SAO – together with GSFC and several commercial spacecraft providers – developed preliminary concepts that concluded that the spacecraft could be built with technologies that are fully mature today – the SSE/SSE+ concept will therefore not push spacecraft capabilities. All subsystems utilize established hardware with substantial flight heritage, including the Ka-band communications system developed for SDO. Most components are "off-the-shelf."

### 3.5 Mission and Science Operations

An initial concept has been developed that describes the mission operations approach and the supporting ground data systems. Key aspects are (1) flight and science operations will be conducted from a joint SSE Science and Operations Center, staffed 5x8; (2) all instruments observe simultaneously; (3) there are daily command uploads (similar to AIA or XRT operations); (4) adequate on-board storage and downlinks avoid the need for on-board flare triggers that have proven difficult to implement. The instrument FOVs are sufficiently large that neither external flare triggers nor repointing will be required (see Fig. 1).

With the data volumes and downlink capabilities listed in Table 2 SSE (or SSE+) can observe continuously 24 hours/day. With the rate listed XMS will be downloading photons at an average rate of 300 photons/pixel/s and HRCI would be taking full 4k×4k images with a cadence of ~1s. The daily data volume is about 40% of SDO's. A typical day of SSE data would be (1) a tracked solar active region with 1s spectra (Fig. 1) at every XMS pixel and 131Å and 193Å images every ~2s, and (2) a full scan of the solar disk at full spectral resolution of the XMS and (3) a full scan of the solar disk at 131Å and 193Å at HRCI resolution, thus creating a rich set of data products that can be used to address a wide array of science problems. These full sun scans will take less than an hour per day.

A wide range of quicklook data products can be produced: 1k×1k (binned) HRCI movies, line-specific (e.g. Fe XVII) images from XMS, light curves in different spectral lines or passbands at each XMS pixel, ion-dependent velocity maps at each XMS pixel, and (with SSE+) hard X-ray light curves from FOXSI.

### 4. MISSION COST AND SCHEDULE

Best-estimate costs for SSE were assembled using various approaches for the different instruments and configurations, including parametric cost modeling, analogy, grass-roots and – given the level of mission maturity – educated estimates. Cost estimates

**Table 3: Instrument Costs and Methodologies**

| Instrument | Methodology | Est. Cost |
|---|---|---|
| XMS | Bottom-up based on detailed WBS and MEL (IXO, Astro-H); analogy (Astro-E, Astro-E2); PRICE-H models (IXO project); PRICE-H (GSFC IDL study) | $100M* |
| HRCI | Analogy (AIA telescopes on SDO; IRIS telescope) | $30M |
| FACTS | parametric; analogy (EIS) | $85M |
| FOXSI | Bottom-up; analogy (NuSTAR) | $40M |

*Does not include $8M for optical bench (carried with S/C in mission costs).*

**Table 4: Costs by WBS**

| | | Estimated Cost ($M, FY10) | | |
|---|---|---|---|---|
| WBS | Description | SSE (Small) | SSE+ (Medium) | RAM (Large) |
| 1,2,3* | PM/SE/SMA | 33 | 42 | 67 |
| 4 | Science | 13 | 20 | 26 |
| 5 | Payload | | | |
| | XMS | 100 | 100 | 100 |
| | HRCI | 30 | 30 | 75** |
| | FOXSI | – | 40 | 40 |
| | FACTS | – | – | 85 |
| 6 | Observatory | 93 | 93 | 93 |
| 7 | Mission Ops | 15 | 17 | 19 |
| 8 | Launch Serv. | 75 | 75 | 75 |
| 9 | GDS | 5 | 6 | 7 |
| 10* | System I&T | 9 | 12 | 20 |
| 11 | EPO | 3 | 4 | 5 |
| Reserves (30% of total)§ | | 89 | 106 | 158 |
| Total | | 465 | 544 | 769 |

\* Wrap factors for WBS1,2,3&10 scale w/complexity
\*\* Ultra-high imager rather the standard HRCI
§ Excludes LV (WBS7) and GO program (in WBS4).

have been generated for every element within the mission Work Breakdown Structure (WBS). For the XMS significant cost credibility exists since it draws heavily on the Astro-H and IXO calorimeter approaches which have been extremely well studied. The EOB and its metrology system are both costed with the S/C. Table 3 summarizes the methodologies used to cost the SSE instruments. These do not include Phase E related costs (i.e., ground system development, operations, etc.) nor science except what is required to design and test the instruments; these other efforts are carried in the appropriate WBS element (see Table 4).

Table 4 summarizes the overall mission cost estimates in the scalable configurations, including Phase E costs. Note that a first order "complexity factor" was accounted for by increasing the wrap percentages for WBS1,2,3 and WBS 10 as the number of instruments increases, and that WBS7 and 9 also increase as the mission scope increases. For a payload complement for SSE+ we assumed FOXSI ($40M) instead of FACTS ($85M), though either spectrometer could be flown and a science-based trade study should be done.

### 4.1 SSE schedule

An overall SSE mission schedule (in its small strategic mission configuration) was generated to determine an NET launch date, and is provided in Fig. 3, assuming a FY13 start. The critical path (which flows through the XMS detector system and then to Observatory I&T) is highlighted on the schedule. Nominal start and delivery dates for all major elements are shown. The schedule supports a September 2018 launch readiness review, with 62 months for development through on-orbit checkout (Phases B, C, and D). This includes a total of 6 months of schedule slack on the critical path which slightly exceeds the NASA guidelines.

From the science requirements perspective, there is no strong driver on SSE/SSE+ relative to the solar cycle. The science objectives of the mission are studies of the quiet and active solar corona, as well as X-ray flares. Regardless of the launch date, the nominal three-year lifetime ensures that an adequate portion of the solar cycle containing numerous flares will be observed.

The schedule reflects the ability to capitalize on the modular nature of SSE. The three main observatory modules (Instrument Module, Spacecraft Module, and EOB Module) will be developed and qualified in parallel, and then delivered for final observatory Integration and Test (I&T). Time has been allocated as appropriate for each of the processes for solicitation, selection, and contract awards. Because the critical path flows through the XMS, and the other instruments being considered for SSE+ and RAM have shorter development times, neither the NET launch date nor the critical path should be heavily affected by adding these elements, though some additional time (~1 month) during Observatory I&T may be required due to the increased complexity.

The modularity of SSE/SSE+ also enables potential cost-sharing with ESA, ESA member states, or JAXA with clearly delineated interfaces, though our cost estimates reflect a NASA-only approach.

## 5. DECADAL EVALUATION CRITERIA

*Identified as a high priority or requirement in previous studies or roadmaps*: The core SSE science has been highly recommended by the 2006 & 2009 Heliophysics roadmaps. SSE is an evolution of the RAM mission (Bookbinder et al. 2003; Golub et al. 2010 White Paper) concept that has been studied by SAO since 2000; RAM was recommended in the 2006 Heliophysics Roadmap.

*Figure 3 – Schedule for SSE mission development with nominal launch in late 2018*

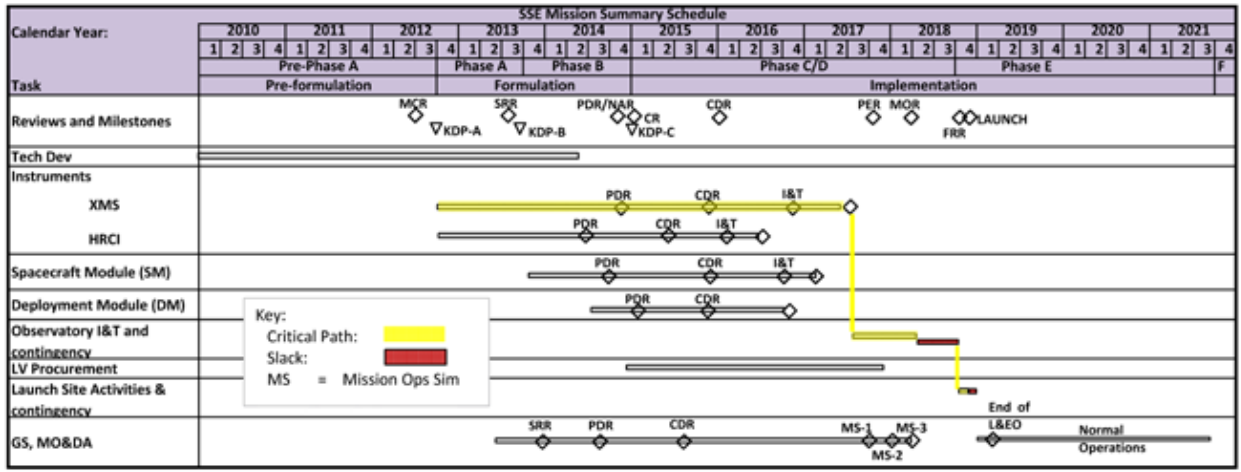

*Makes a significant contribution to more than one of the Panel themes*: SSE is primarily a basic science mission. The rich SSE mission data products (images, spectra, and timing data) will motivate and challenge researchers in theory, modeling and laboratory plasma physics and in atomic physics.

*Contributes to important scientific questions facing solar and space physics today*: The fundamental question in space physics is the nature of instabilities in magnetized plasmas. The storage and release of energy in the plasma is controlled by the onset of the instabilities. The SSE X-ray spectra give the definitive measurement of the thermal state of the plasma as a function of space and time, while the SSE HRCI shows the magnetic configuration that confines, heats and moves the plasma.

*Complements other observational systems or programs available*: SSE/SSE+ is highly complementary to other planned and conceived space missions with its balanced emphasis between coronal spectroscopy and imaging. When SSE is operational several significant ground-based facilities will be online. Foremost is the Advance Technology Solar Telescope (ATST), whose 4m telescope will primarily study plasma processes in the photosphere and chromosphere. The magnetic flux that creates and defines the solar corona necessarily passes through these high beta (and beta~1) regions. ATST and SSE are complementary in their spectroscopic capabilities. The Frequency Agile Solar Radio Telescope (FASR; Gary et al. 2010) directly images the evolution of coronal electrons in flares and coronal mass ejections, while SSE will measure the properties of the ionic plasma (see Fig. 1), providing a natural complementarity for shock and flare studies. COSMO is an innovative program to measure the magnetic field directly in the corona above the solar limb, used to quantifying the energetics of CMEs.

*Has an appropriate degree of readiness (technical, resources, people)*: The only element of the SSE that requires development is the XMS (see §3 of Bandler et al. 2010b). The technology development plan for the XMS has it achieving TRL5 by 2013, well ahead of an SSE new start.

*Fits with other national and international plans and activities*: The SSE mission will provide the observations necessary to address fundamental science questions outlined by the 2009 Heliophysics Roadmap. SSE will provide complementary science to the heliospheric and solar missions Solar Probe Plus and (potentially) Solar Orbiter (ESA) and Solar-C (JAXA). SSE will give the most comprehensive description of the coronal plasmas' thermal state along field lines that pass near Solar Probe Plus. The knowledge of the state of the plasma at the sun is critical for interpreting the evolution of the plasma as it enters the heliosphere. SSE and Solar Orbiter will simultaneously observe the corona from different view points. Solar-C (Plan B) is a high resolution mission focused on the chromosphere/corona interface. SSE and SSE+ provide complementary coronal observations to Solar-C extending the full complement of spectroscopy from the UV - chromosphere through the soft and hard X-ray corona.

# 6. REFERENCES

*Note: White papers submitted to the heliophysics decadal survey "A Decadal Strategy for Solar and Space Physics" - Solar and Heliospheric Physics panel identified in italics.*

# 7. ACRONYMS

*ADAM – Able Deployable Articulated Mast*
*AIA – Atmospheric Imaging Assembly*
*ADR – Adiabatic Demagnetization Refrigerator*
*CBE – Current Best Estimate*
*CCD – Charge Coupled Device*
*EOB – Extendable Optical Bench*
*ESA – European Space Agency*
*EUV – Extreme UltraViolet*
*FACTS – Fine scale Advanced Coronal and Transition Region Spectrograph*
*FMA – Flight Mirror Assembly*
*FOV – Field of View*
*FOXSI – Focusing Optics X-ray Solar Imager*
*GI – Grazing Incidence*
*HRCI – High Resolution Coronal Imager*
*IXO – International X-ray Observatory*
*NET – No Earlier Than*
*NuSTAR – Nuclear Spectroscopic Telescope Array*
*RAM – Reconnection And Microscale*
*S/C – SpaceCraft*
*SMEX – Small Explorer*
*SR&T – Supporting Research & Technology*
*SSE – Solar Spectroscopy Explorer*
*SSE+ – SSE with XMS, HRCI, and FOXSI*
*SSOC – SSE Science and Operations Center*
*TRACE – Transition Region and Coronal Explorer*
*TRL – Technology Readiness Level*
*WBS – Work Breakdown Structure*
*XMS – X-ray Microcalorimeter Spectrometer*